\documentstyle[eqsecnum,aps,prd]{revtex}

\begin{document}
\thispagestyle{empty}
{\baselineskip0pt
\leftline{\large\baselineskip16pt\sl\vbox to0pt{\hbox{DAMTP} 
               \hbox{University of Cambridge}\vss}}
\rightline{\large\baselineskip16pt\rm\vbox to20pt{
               \hbox{DAMTP-1999-24}
               \hbox{UTAP-324}
               \hbox{RESCEU-5/99}
               \hbox{\today}
\vss}}%
}
\vskip15mm

\begin{center}
{\large\bf Gravitational Mass in Asymptotically de Sitter 
Space-Times with Compactified Dimensions}
\end{center}

\begin{center}
{\large Tetsuya Shiromizu 
\footnote{JSPS Postdoctal Fellowship for Research Abroad}} \\
\vskip 3mm
\sl{DAMTP, University of Cambridge \\ 
Silver Street, Cambridge CB3 9EW, UK \\
\vskip 5mm
Department of Physics, The University of Tokyo, Tokyo 113-0033, 
Japan \\
and \\
Research Centre for the Early Universe(RESCEU), \\ 
The University of Tokyo, Tokyo 113-0033, Japan
}
\end{center}
\vskip 5mm
\begin{center}
{\it to be published in Phys. Rev. D}
\end{center}
\begin{abstract}
We define gravitational mass in asymptotically de Sitter space-times with 
compactified dimension. It was shown that the mass can be
negative for space-time with matter spreading beyond the cosmological horizon 
scale or large outward `momentum' in four dimension. We 
give simple examples with negative energy in higher dimensions even 
if the matter is not beyond horizon or system does not have large `momentum'. 
They do not have the lower bound on the mass. We also give 
a positive energy argument in higher dimensions and realise that 
elementary fermion cannot exist in our examples. 
\end{abstract}

\section{Introduction}


Superstring or M-Theory may offer the proper theory of gravity
\cite{String}. Suck kind of   
theories are formulated in higher dimensions and it is believed that 
the extra dimension space will be compactified to be 
less than Planck length. 

The stability of such space-times is important and has been 
discussed. It was shown that there is 
the instanton which may indicate the decay of 
Kaluza-Klein vacuum\cite{Witten}\cite{Brill1}\cite{Brill2}. As Witten
pointed out, the decay mode is excluded by the existence of 
(massless) elementary fermion related to supersymmetry\cite{Marika}. 

On the other hand, the stability of the asymptotically anti-de Sitter(AdS) 
space-time with compactified dimension has been focused recently
by Horowitz \& Myers\cite{Myers} 
because AdS/CFT correspondence\cite{AdS} links 
the stability of super-Yang-Mills theory to that of AdS space-time. So they 
suggested a positive energy conjecture in locally asymptotically 
AdS space-times. 

In the actual cosmological context, same argument of stability is 
important. In this paper, for simplicity, we consider D-dimensional Einstein
gravity with a positive cosmological constant. A positive 
cosmological constant is essential for inflation
universe\cite{Inflation} 
and acceleration of universe confirmed gradually by recent 
observations of supernovae\cite{SN}. 
In Nakao et al's study, it was shown that the 
gravitational mass can be negative in four dimensional space-times
when the matter distributes beyond the cosmological horizon
scale\cite{Nakao}\cite{Nakao3}.  
In other words, the mass is negative if the space-time has a  
large outward `momentum'. Evaluating the electric part of Weyl tensor, they 
also checked that the mass is related to gravitational tidal force. 
In this paper we find a 
initial data set whose mass can be negative and does not have 
the lower bound even if the space-time does not have
`momentum'. Furthermore, we give an other dynamical solution which 
also does not have `momentum'. 

In asymptotically flat space-times with compactified extra dimensions,
there are examples of momentarily static initial slices such that the 
energy can be negative regardless of the size of the compactified 
dimensions\cite{Brill2}. A parallel argument helps us to discuss 
about the lower bound of the energy in asymptotically de Sitter
space-times. Since a positive cosmological constant prompts 
the rapid expansion of the universe, we might be able to
obtain an implication about 
the effect to the energy from the dynamics of the compactified
dimensions. 

The rest of the present paper is organised as follows. In the next
section we give the definition of gravitational mass in
asymptotically de Sitter space-time with extra dimensions and write 
down the expression in terms 
of canonical quantities on the hypersurfaces. Since a part of this is 
straightforward extension of work by Nakao et
al\cite{Nakao} based on \cite{AD}\cite{Deser}, we give the brief discussion. 
We will point out that the expression is a special case of the gravitational 
Hamiltonian defined by Hawking \& Horowitz\cite{HH}. In Sec. 3 we 
show two examples which have negative energy and no naked singularity. In
Sec. 4, we consider the positive energy theorem to discuss the stability 
of de Sitter space-time. As a result we confirm 
that elementary fermion cannot exist in examples given by us. 
Finally we give summary in Sec. 4.

\section{Gravitational Mass in Asymptotically 
de Sitter Space-Times with Higher Dimensions}

We consider D-dimensional space-times which satisfy the 
Einstein equation with a positive cosmological constant $\Lambda$,
%
\begin{eqnarray}
R_{IJ}-\frac{1}{2}g_{IJ}R+\Lambda g_{IJ}=8 \pi G_D T_{IJ},
\end{eqnarray}
%
where suffices $I,J$ runs over $0, 1, ..., D-1$, $T_{IJ}$ is the 
energy-momentum tensor and $G_D$ is D-dimensional Newton's constant. 
We decompose the space-time metric into the D-dimensional de Sitter
metric ${\bar g}_{IJ}$ and the rest$h_{IJ}$; 
%
\begin{eqnarray}
g_{IJ}={\bar g}_{IJ}+h_{IJ}.
\end{eqnarray}
%
Hereafter the notation `over-bar' 
indicates the quantities of the background de Sitter space-time. 
We remember that $h_{IJ}$ 
is not necessarily small, but we impose that it vanishes at infinity. 
We will give a further description below.

Here we note that Einstein equation is written as 
%
\begin{eqnarray}
R_\ell^{IJ}-\frac{1}{2}{\bar g}^{IJ}R_\ell -\Lambda h^{IJ}
=(-{\bar g})^{-1/2}{\cal T}^{IJ}, \label{eq:linear}
\end{eqnarray}
%
where $h^{IJ}={\bar g}^{IK}{\bar g}^{JL}h_{KL}$, 
$R^{IJ}_\ell$ is the linear part of Ricci tensor $R_{IJ}$ with respect
to $h_{IJ}$ 
and all the higher order terms are included in ${\cal T}^{IJ}$ of 
the right-hand side.  
As the left-hand side of eq. (\ref{eq:linear}) 
satisfies the Bianchi identity, we see that 
%
\begin{eqnarray}
{\bar \nabla}_I {\cal T}^{IJ} =0
\end{eqnarray}
%
holds, where ${\bar \nabla}_I$ is the covariant derivative with
respect to ${\bar g}_{IJ}$.  
Contracting ${\cal T}^{IJ}$ with the Killing vector of de Sitter space-time,
%
\begin{eqnarray}
{\cal T}^I={\cal T}^{IJ}{\bar \xi}_J,
\end{eqnarray}
%
we obtain the local conservation low 
%
\begin{eqnarray}
\partial_I {\cal T}^I =0. 
\end{eqnarray}
%
Thus, $ \int d^{D-1}x {\cal T}^0$ is conserved if the surface term 
$ \int d^{D-2}S_i{\cal T}^i$ vanishes. We define  
%
\begin{eqnarray}
E_{\rm AD} & = & 
\frac{1}{8\pi G_D}\int d^{D-1}x {\cal T}^{0K}{\bar \xi}_K \nonumber \\
& =& \frac{1}{8\pi G_D} \int d^{D-2}S_i [{\bar \nabla}_I
K^{0iJM}-K^{0jJi}{\bar \nabla}_j ]{\bar \xi}_J, \label{eq:First}
\end{eqnarray}
%
where $K^{IJKL}=(1/2)({\bar g}^{IL}H^{KJ}+{\bar g}^{KJ}H^{IL}
-{\bar g}^{IK}H^{JL}-{\bar g}^{JL}H^{IK})$, $H^{IJ}=h^{IJ}
-(1/2){\bar g}^{IJ}h^K_K$ and $i=1, 2, ..., D-1$. In the case that 
${\bar \xi}^I$ is a timelike Killing vector $E_{\rm AD}$ is 
regarded as the Killing energy, the so called Abbott-Deser(AD) mass. 

Next, we rewrite eq. (\ref{eq:First}) in more familiar form 
in order to obtain its physical meaning. First of all, we need take 
asymptotic region carefully. In asymptotic flat cases, the 
extrinsic curvature $K_{ij}$ of slices has the 
behaviour $ K^i_j \rightarrow 0$ 
toward the spatial infinity $i^0$. As we know, 
we can expect that such slice does not have asymptotic region in 
asymptotically de Sitter space-time because static slices with 
$K_{ij}=0$ have no boundary in de Sitter space-time. So we remember 
that the flat chart of de Sitter space-time has boundary and 
its spatial metric is conformally flat(See Fig. 2 in ref
\cite{Nakao}). 
Thus, we realize that the most natural 
condition is given by $K^i_j \rightarrow H \delta^i_j $ or 
$K \rightarrow (D-1)H$ \cite{Nakao2}, where $H={\sqrt
{2\Lambda/(D-1)(D-2)}}$.  
In this chart the metric of the background de Sitter space-time is written as 
%
\begin{eqnarray}
{\bar g}=-dt^2+a(t)^2\sum_{i=1}^{D-1} dx_i^2,
\end{eqnarray}
%
where $a(t)=e^{Ht}$. The Killing vector has 
the component ${\bar \xi}^M=(-1, Hx^i)$. After simple calculations we 
obtain
%
\begin{eqnarray}
E_{AD} & = & 
a(t) \Bigl( E_{\rm ADM}+\Delta P_{\rm ADM}(-{\bar \xi}) \Bigr)\nonumber \\
& = & \frac{a(t)}{16 \pi G_D}\int d{\bar S}_i(\partial_jh^{ij}-{\bar g}^{ij}
\partial_j h^k_k) -\frac{a(t)}{8\pi G_D} \int d{\bar S}_i 
\Bigl[ K^i_j-K\delta^i_j+(D-2)H \delta^i_j \Bigr] {\bar \xi}^j,\label{eq:AD}
\end{eqnarray}
%
where $K_{ij}$ is the extrinsic curvature of $t=$constant 
$(D-1)$-dimensional hypersurface. By using the momentum 
$\pi_{ij}=K_{ij}-q_{ij}K$, where $q_{ij}$ is the metric of 
$t=$constant hypersurface, the second term in the second line 
of eq. (\ref{eq:AD}) can be written as  
%
\begin{eqnarray}
\Delta P_{\rm ADM}(-{\bar \xi})=P_{\rm ADM}(-{\bar \xi})-
{\bar P}_{\rm ADM}(-{\bar \xi})=-\frac{1}{8\pi G_D}\int d{\bar S}_i
( \pi^i_j-{\bar \pi}^i_j ) {\bar \xi}^j.  
\end{eqnarray}
%
%
%
As a result, $E_{\rm AD}$ is written in terms of the sum of the ADM energy 
and net momentum. We note that the momentum of background de Sitter 
space-time is subtracted automatically. The above net momentum is the 
term `momentum' used in introduction. The above argument does not 
depends on whether extra dimensions is compactified or not.  

Now we compare the above expression with the gravitational energy 
defined via `physical Hamiltonian'\cite{HH}. The gravitational energy 
 has expression  
%
\begin{eqnarray}
E_{\rm HH}=-\frac{1}{8\pi G_D}\int dS(Nk-{\bar N} {\bar k})
+\frac{1}{8\pi G_D}\int dS(N^i \pi_{ij}
-{\bar N}^i{\bar \pi}_{ij})r^j, \label{eq:HH}
\end{eqnarray}
%
where $k$ and $r^i$ are the trace of the extrinsic curvature and 
the unit normal vector of $(D-2)$-dimensional surface at the infinity,
respectively. $N$ and $N^i$ are the lapse function and the shift vector. 
  
The `physical Hamiltonian' is defined by the substraction the Hamiltonian 
of a background space-time from the original Hamiltonian. 
As Hawking \& Horowitz showed,
the first term of eq. (\ref{eq:HH}) is just the ADM energy. 
Comparing eq. (\ref{eq:AD}) with eq. (\ref{eq:HH}), 
one can see that they have the same expression 
except for scale factor, {\it i.e}, $E_{\rm AD}=a(t)E_{\rm HH}$, 
if one chooses the lapse function and the shift vector as follows, 
%
\begin{eqnarray}
N={\bar N}=1 ~~~~~~~{\rm and}~~~~~~~N^i={\bar N}^i=-{\bar \xi}^i. 
\end{eqnarray}
%
%
%
It is likely that the priority is given to our definition 
in asymptotically de Sitter space-times because the 
Hawking \& Horowitz's construction of physical Hamiltonian has 
the operation of artificial substraction to keep it finite.  

In asymptotically flat cases, the ADM energy is defined 
by view point of a static observer with $N=1$ and $N^i=0$ at 
spatial infinity $i^0$. The naturalness of observer selection to 
define the energy is related to the timelike translation symmetry 
at infinity. On the other hand, in asymptotically de Sitter space-time, we 
must consider hypersurfaces which have infinity in order to 
define the non-zero energy. As we have done, the most 
convenient slices is one which corresponds to the flat slices 
in de Sitter space-time. This slices reach into the timelike infinity 
${\cal I}^+$(See Fig. 2 in Ref.\cite{Nakao}). 
However, the slices is not associated with the timelike 
translation symmetry which de Sitter space-time possesses, that is, 
the slice is not orthogonal to timelike Killing vector. Thus, it 
may be natural that the ADM momentum term enters into the expression 
of the mass. 

In four dimension the AD mass can be negative when system has a large outward 
momentum\cite{Nakao}. On the other hand, the positivity of the AD or ADM mass 
is guaranteed for systems without the net momentum, 
$\Delta P_{\rm ADM}=0$, in four dimension\cite{Tess}.   

\section{Examples in five dimensions}

In the same way as four dimensional cases, the 
AD mass can be negative regardless of 
the compactification of the extra dimension if the system has a large
momentum. However, the 
physical reason of the 
negativity is the dynamics of four dimensional part of space-time, 
rather than the (quantum)
stability. Now we are 
interest in just stability. So, we consider the situation only 
where the contribution 
of the net ADM momentum does not exist. We give two examples in five
dimension which 
are regular everywhere and have negative energy. 

Let us consider an initial slice with $K_{ij}=\pm Hg_{ij}$ and 
$H={\sqrt {\Lambda/6}}$. In this slice the Hamiltonian constraint becomes   
$ {}^{(4)}R=0$. Thus one can use the argument on the momentarily static
slices in asymptotically flat cases because the Hamiltonian constraint 
is just same one. 

One can see easily that the euclidian Reissner-Nordstrom 
metric with imaginary `charge' $ie$ satisfies the Hamiltonian 
constraint\cite{Brill2}. This metric of the hypersurface is given by     
%
\begin{eqnarray}
{}^{(4)}g=V(r)d\chi^2+\frac{dr^2}{V(r)}+r^2 d\Omega_2^2
\end{eqnarray}
%
where $V(r)=1-2m/r-e^2/r^2$ and $r \geq r_+:=
m+{\sqrt {m^2+e^2}}$. To avoid a conical singularity at 
$r=r_+$, we assume the period $\chi_p=4\pi/V'(r_+)=2\pi r_+^2/(r_+-m)$ 
along the $\chi$-direction. 

On the present slice, the mass is constructed by only the ADM energy 
component, $E_{\rm AD}=E_{\rm ADM}=m\chi_p/2G_5=m/2$\footnote{Here we used 
the fact that the five dimensional Newton's constant is written as $G_5=
\chi_p G_4$ in term of the four dimensional one.}.  
By the same argument as Brill \& Horowitz\cite{Brill2}, it is shown 
that the mass becomes
negative and does not have the lower bound. The mass can be set to be 
arbitrary negative regardless of the radius of the compactified
space. However, it is not obvious 
whether the AD energy is conserved or not in the course of time
development because of the existence of the cosmological
constant and momentarily non-static slices. In fact we will see that 
the energy is not conserved in the next example. 

As a second example, we take a dynamical solution\cite{Maki}. The metric is 
%
\begin{eqnarray}
ds^2=-dt^2+a(t)^2\Bigl[ \frac{d \chi^2}{\Delta}+\Delta dr^2+r^2 \Delta^2 
d\Omega_2^2\Bigr],
\end{eqnarray}
%
where $\Delta (r)=1-m/r$, $a(t)=e^{\pm Ht}$ and $H={\sqrt {\Lambda/6}}$. 
The $t=$constant hypersurface has the 
extrinsic curvature $K^i_j=\pm H \delta^i_j$. However, this space-time
has 
timelike naked singularity at $r=m$. Since the expansion of outgoing 
null geodesics congruence is 
%
\begin{eqnarray}
\theta_+=\frac{1}{{\sqrt {2}}}\Bigl(\pm
2H+\frac{4r-m}{2a\Delta^{3/2}r^2} \Bigr),
\end{eqnarray}
%
the apparent horizon does not exist in the expanding chart with $a=e^{Ht}$. 
In collapsing chart with $a=e^{-Ht}$, 
apparent horizon also does not exist although surfaces such that 
$\theta_+=0$ exists. The curvature invariant is given by 
%
\begin{eqnarray}
R_{IJKL}R^{IJKL}=5H^4+3\Bigl[H^2-\frac{m}{2a^2(r-m)^3}
\Bigr]^2+\Bigl[H^2-\frac{m}{a^2(r-m)^3} \Bigr]^2.
\end{eqnarray}
%

To avoid the naked singularity at $r=m$, 
we change the sign of the mass parameter, $m \rightarrow -m$. After
that the radial coordinate $r$ can run up to $r=0$ and the conical
singularity occurs at $r=0$ in general cases. Near $r=0$, the metric 
is written as 
%
\begin{eqnarray}
ds^2 \simeq -dt^2+a(t)^2 \Bigl[ \frac{r}{m}d\chi^2+\frac{m}{r} dr^2 
+m^2d\Omega^2_2 \Bigr].
\end{eqnarray}
%
Here we introduce a new coordinate $R = (rm)^{1/2}$ and the metric is 
%
\begin{eqnarray}
ds^2 \simeq -dt^2+ a^2 \Bigl[4\Bigl\lbrace dR^2+R^2 d \Bigl(
\frac{\chi}{2m}\Bigr)^2 \Bigr\rbrace +m^2d\Omega^2_2\Bigr].
\end{eqnarray}
%
Hence, the metric is regular everywhere except for $a=0$ singularity 
if one assumes the period $\chi_p=4\pi m$ along the $\chi$-direction. 
The physical size of the
compactified dimensions is given by  $4 \pi a(t) m$. Thus, it 
decreases/increases if one takes the collapsing/expanding chart. 
A cosmological constant make the compactified space dynamical as well as 
the four dimensional part. The AD mass is  
%
\begin{eqnarray}
E=E_{\rm ADM}=-\frac{a(t)^3m \chi_p}{G_5}=-a^3(t)m. 
\end{eqnarray}
%
This is not conserved due to the expansion of the universe! 
This comes from the non-vanishing boundary
term $\int d^{D-2}S_i{\cal T}^i$ which vanishes for asymptotically flat cases. 
At first glance, best we can do might be choosing the collapsing chart and 
we can keep the radius of the compactified space to be 
less than Planck length. In the chart, we observe the value of the 
energy approaches zero. One may think that a large cosmological
constant stabilise the space-times because 
the extra dimension shrinks rapidly and the energy becomes zero before 
space-time decays. At the same time the shrinking, however, means the 
big-crunch of the space-time and then this case does not give
attractive model. Since there is no reason why one has to 
choose the collapsing chart {\it a priori}, we need consider the
expanding chart too. In this chart, the compactified space expands
till the end of the inflation and the the absolute value of the AD
energy increases. Naively speaking, the decay rate of the de Sitter 
space-time into the space-time with the large compactified space is 
suppressed.

\section{Positive Energy Theorem, Elementary Fermion and Stability}

In locally asymptotically flat cases with compactified dimensions, 
the break down of positive energy
theorem means non-existence of the Witten spinor
\cite{Marika}. In supergravity side, that means there are not supersymmetry
because the spinor is related to infinitesimal generator of local 
supersymmetry\cite{Gary}. 

In asymptotically de Sitter cases, the situation is different from the 
above. As we stated, the energy can be negative even if the extra 
dimension is not compactified and the topology is trivial. 
In this section, we discuss the positive 
energy theorem, based on \cite{Tess}\cite{Kastor}
\footnote{The argument in ref. \cite{Tess} is prototype of the proof. 
Rigourously speaking, the `supercovariant derivative' defined there is not 
covariant for full space-times. 
The excellent approach bearing supergravity in mind was given in ref. 
\cite{Kastor}. However, 
both papers have not been attained in the present refined form. The 
explicit evaluation of the left-hand side of eq. (\ref{eq:Witten}) has
not done yet in ref. \cite{Kastor}. 
See ref. \cite{Gibbons} for asymptotically AdS space-times.}, in cases  
whose extra dimension is not compactified. Conversely, we can see
easily that the Witten spinor cannot exist for 
examples given in the previous section. 

Following Kastor\&Traschen\cite{Kastor}, we define the cosmological 
supercovariant derivative operator on a spinor $\epsilon$ as 
%
\begin{eqnarray}
{\hat \nabla}_I \epsilon
=  \Bigl(\nabla_I+\frac{i}{2}H\gamma_I\Bigr)\epsilon =
\Bigl( \partial_I+\Gamma_I +\frac{i}{2}H \gamma_I \Bigr) \epsilon
=:\Bigl(\partial_I + \Gamma'_I \Bigr)\epsilon,
\end{eqnarray}
%
where $\Gamma_I$ is the spin connection, 
%
\begin{eqnarray}
\Gamma_I=-\frac{1}{8}e^{J {\hat K}} \nabla_I e_J^{{\hat L}}
[\gamma_{{\hat L}}, \gamma_{{\hat K}}], 
\end{eqnarray}
%
and $e^I_{\hat J}$ is quasi-orthogonal basis. The cosmological 
Witten equation\footnote{ The vector $\xi^I = {\bar \epsilon}\gamma^I 
\epsilon $ defined by spinor $\epsilon$ satisfying ${\hat \nabla}_I
\epsilon=0$ is conformal Killing vector, not Killing vector in 
de Sitter space-time. That it, $\epsilon$ is a conformal 
Killing spinor, not Killing spinor. 
This point is the main reason why one cannot 
prove the positivity of the AD mass for asymptotically de Sitter
space-times having the net momentum because the left-hand side 
of eq. (\ref{eq:Witten}) cannot be the AD mass.} is define by 
%
\begin{eqnarray}
\gamma^i {\hat \nabla}_i \epsilon = 0. 
\end{eqnarray}
%
The solution is given by a constant 
spinor $\epsilon_0$ satisfying $\gamma^{\hat
0}\epsilon_0=-i\epsilon_0$ in the expanding flat slice of de Sitter 
space-time. 

By using the Bianchi identity, the Hamiltonian and momentum constraints, 
we obtain an identity
%
\begin{eqnarray}
\int dS^i \epsilon^\dagger {\hat \nabla}_i \epsilon
=\int dV\Bigl[ |{\hat \nabla}_i\epsilon|^2+4 \pi G_D (\epsilon^\dagger
T_{{\hat 0}{\hat 0}} \epsilon +\epsilon^\dagger T_{i{\hat 0}}\gamma^i 
\gamma^{\hat 0}\epsilon) \Bigr] \label{eq:Witten}
\end{eqnarray}
%
Let us evaluate the left-hand side. We follow Witten's
argument\cite{Edward} carefully. First, the constant spinor satisfies 
%
\begin{eqnarray}
\gamma^i{\hat \nabla}_i \epsilon_0=\gamma^i \Gamma'_i \epsilon_0
=\gamma^i {}^{(D-1)}\Gamma_i \epsilon_0 +\frac{i}{2}[-K+(D-1)H]\epsilon_0
\end{eqnarray}
%
where ${}^{(D-1)}\Gamma_i=-(1/8)e^{j{\hat k}}{}^{(D-1)}\nabla_i 
e_k^{\hat \ell}[\gamma_{\hat \ell}, \gamma_{\hat k}] $ and we used 
%
\begin{eqnarray}
\Gamma_i'={}^{(D-1)}\Gamma_i+\frac{1}{2}K_{ij}\gamma^i\gamma^{\hat 0}+
\frac{i}{2}H\gamma_i  \label{eq:DC}
\end{eqnarray}
%
If we suppose $K=(D-1)H+O(1/r^{D-1})$, the 
same argument of the existence of solution as asymptotic flat holds 
and we obtain   
%
\begin{eqnarray}
\int dS^i \epsilon^\dagger {\hat \nabla}_i \epsilon
=\int dS^i {\epsilon}_0^\dagger
(\Gamma'_i-\gamma_i \gamma^j \Gamma'_j){\epsilon}_0.
\end{eqnarray}
%
Inserting the decomposition (\ref{eq:DC}) of the spin connection into 
the above, we obtain the 
familiar expression 
%
\begin{eqnarray}
\int dS^i \epsilon^\dagger {\hat \nabla}_i \epsilon
& = & \int dS^i {\epsilon}_0^\dagger
({}^{(D-1)}\Gamma_i-\gamma_i \gamma^j {}^{(D-1)}\Gamma_j){\epsilon}_0
+\frac{1}{2}\int dS_i {\epsilon}_0^\dagger 
\Bigl[ K^i_j-\delta^i_jK+(D-2)H\delta^i_j \Bigr] \gamma^j
\gamma^{\hat 0}{\epsilon}_0 \nonumber \\
& = & \frac{1}{4}\int dS^i{\epsilon}_0^\dagger(\partial_j h_i^j-
\partial_ih^j_j){\epsilon}_0-
\frac{1}{2}\int dS_i  {\tilde K}^i_j 
{\bar \epsilon}_0 \gamma^j {\epsilon}_0 \nonumber \\
& = & 4\pi G_D \Bigl(  E_{\rm ADM}|\epsilon_0 |^2+\Delta P_{\rm ADM}
({\bar \epsilon}_0 {\bf \gamma}\epsilon_0 ) \Bigr), 
\end{eqnarray}
%
where ${\tilde K}^i_j$ is the traceless part of $K^i_j$. 
Here we note that ${\bar \epsilon}_0 \gamma^i \epsilon_0=-e^i_{\hat 0}
|\epsilon_0|^2 = O(1/r^{D-3})$ due to $\gamma^{\hat 0}\epsilon_0=
-i \epsilon_0$. This and ${\tilde K}^i_j=O(1/r^{D-2})$ lead us that 
the net momentum term $\Delta P_{\rm ADM}({\bar \epsilon}_0 {\bf
\gamma}\epsilon_0 )$ vanishes. Finally we obtain an inequality
%
\begin{eqnarray}
E& = & \frac{a(t)}{4\pi G_D} \int dS^i \epsilon^\dagger 
{\hat \nabla}_i \epsilon 
\nonumber \\
& = & a(t)E_{\rm ADM}|\epsilon_0 |^2 \geq 0 \label{eq:positive}
\end{eqnarray}
%
under the dominant energy condition\cite{Wald} on the energy-momentum tensor
$T_{IJ}$. Unfortunately, we cannot say something about $E=0$ case because 
Witten spinor approaching to $\epsilon_0$ at infinity is uniquely 
determined\footnote{In asymptotically flat space-times, $\epsilon_0$ 
is arbitrary constant spinor. This means the existence of 
D's independent solution of Witten equation. As a result, we obtain 
$R_{IJKL}=0$ from $0=[\nabla_i, \nabla_j]\epsilon =(1/4)R_{ijKL}
[\gamma^K,\gamma^L] \epsilon $.}. 
Thus, there is not contradiction with the existence of 
non-trivial solution which has zero ADM mass and is 
given in ref. \cite{Nakao}. 

We can see that $E$ is not equal to the AD mass, $E_{\rm AD}$. When ${\tilde 
K}^i_j=O(1/r^{D-1})$ holds and the momentum term of the AD mass vanishes, 
the AD mass equals to the ADM energy and the positivity is guaranteed. 
For our purpose, it is worth imposing ${\tilde K}^i_j=O(1/r^{D-1})$ 
because we are interest in the stability of space-time, not its dynamics. 
As eq. (\ref{eq:positive}) this implies the positivity of the AD 
mass. The apparent contradiction with examples given in the previous 
section indicates that the Witten spinor does not exist in 
such examples.

\section{Summary}

In this paper we defined the gravitational mass in asymptotically 
de Sitter space-time with extra dimensions and obtained a refined  
expression related to the ADM energy and momentum associated to the timelike 
Killing vector of the background de Sitter space-time. 
Furthermore, we gave one dynamical solution and one 
initial data with the negative energy in five dimensions. Does 
these solution indicate the quantum decay of de Sitter space-time? 
We cannot reply to the question instantly because we do not know whether 
the instanton exists or not. Naively speaking, we can guess from 
the previous section that the decay occurs unless one imposes the 
existence of elementary fermion or supersymmetry. 

Finally, we should stress the fact that the energy seems to depend on
the time in general although the definition is reasonable. This means 
that the contribution from the boundary is not negligible. 

\section*{Acknowledgements}

The author is grateful to Gary Gibbons for discussion in the early
stage of this work and DAMTP relativity group for their hospitality. 
He also thanks M. Spicci for his careful reading of this manuscript. 
This work is supported by JSPS fellowship(No. 310).

\end{document}